\documentclass[iop,apjl]{emulateapj}
\usepackage{amsmath}
\usepackage{graphicx}
\usepackage{natbib}
\usepackage{color}

\usepackage[colorlinks,citecolor=blue,linkcolor=red]{hyperref}
\usepackage{longtable}
\usepackage{threeparttable}
\usepackage{rotating}
\usepackage{multirow,booktabs}
\usepackage{wasysym}
\usepackage{float}

\newcommand{\lum}{erg~s\ensuremath{^{-1}}}

\begin{document}

\bibliographystyle{apj} 
\title{Initial results from a systematic search for changing-look active galactic nuclei selected via mid-infrared variability}

\author{Zhenfeng Sheng\altaffilmark{1,2$\dag$}, Tinggui Wang\altaffilmark{1,2$\ddag$}, Ning Jiang\altaffilmark{1,2},Jiani Ding\altaffilmark{3}, Zheng Cai\altaffilmark{4}, Hengxiao Guo\altaffilmark{5,6}, Luming Sun\altaffilmark{1,2}, Liming Dou\altaffilmark{7}, Chenwei Yang\altaffilmark{8}}

\altaffiltext{1}{CAS Key Laboratory for Researches in Galaxies and Cosmology, University of Sciences and Technology of China, Hefei, Anhui 230026, China; $^{\dag}$shengzf@mail.ustc.edu.cn; $^{\ddag}$twang@ustc.edu.cn}

\altaffiltext{2}{School of Astronomy and Space Science, University of Science and Technology of China, Hefei 230026, China}
\altaffiltext{3}{Department of Astronomy \& Astrophysics, University of California Santa Cruz, 1156 High Street, Santa Cruz, CA 95060, USA}
\altaffiltext{4}{UCO/Lick Observatory, University of California, 1156 High Street, Santa Cruz, CA 95064, USA} 
\altaffiltext{5}{Department of Astronomy, University of Illinois at Urbana-Champaign, Urbana, IL 61801, USA} 
\altaffiltext{6}{National Center for Supercomputing Applications, University of Illinois at Urbana-Champaign, 605 East Springfield Avenue, Champaign, IL 61820, USA}
\altaffiltext{7}{Center for Astrophysics, Guangzhou University, Guangzhou 510006, China}
\altaffiltext{8}{SOA Key Laboratory for Polar Science, Polar Research Institute of China, 451 Jinqiao Road, Shanghai, 200136, China}

\begin{abstract}
Changing look active-galatic-nuclei (CL AGNs) can yield considerable insight into accretion physics
as well as the co-evolution of black holes and their host galaxies. A large sample of these CL AGNs
is essential to achieve the latter goal.  We propose an effective method to select CL candidates from  
spectroscopic quasar catalogs using the mid-infrared (MIR) variability information derived from 
\emph{ALLWISE/NEOWISE} data releases. Our primary selection criteria include both a large amplitude flux 
variation and a transition of MIR color from an AGN to a normal galaxy. A pilot spectroscopic follow-up 
of 7 candidates among about 300 candidates selected from Sloan Digital Sky Survey low-redshift (z$<$0.5) 
AGN sample results in 6 new turn-off CL AGNs. We expect to obtain hundreds of CL AGNs once full 
spectroscopic follow-up of the sample is carried out.

\end{abstract}

\section{Introduction}
Changing look active-galatic-nuclei (CL AGNs) are sources which can exhibit a transit from Type 1 to Type 1.8, 1.9 and 2 or vice versa, featuring disappearing or emerging broad emission lines (BELs) on timescales of months to years (LaMassa et al. \citeyearpar{LaMassa2015}; Runnoe et al. (\citeyear{Runnoe2016};  Ruan et al. \citeyear{Ruan2016}; Macleod et al. (\citeyear{MacLeod2016}); Gezari et al. (\citeyear{Gezari2017})).
These sources were once serendipitously discovered in nearby AGNs (e.g.Cohen et al. \citeyear{Cohen1986}; Storchi-Bergmann et al. 1993; Eracleous \& Halpern 2001 ;Denney et al. \citeyear{Denney2014}; Shappee et al. \citeyear{Shappee2014}).
With the repeated spectroscopic observations, the number of CL AGNs has increased rapidly in recent years (Ruan et al. \citeyear{Ruan2016}; MacLeod et al. \citeyear{MacLeod2016},\citeyear{MacLeod2019}; Yang et al. \citeyear{Yang2018}; Oknyansky et al. \citeyear{Oknyansky2019}). 

The origin of CL AGN is still under debated. Between the two main possible explanations, the rapid changes in black hole accretion rate is preferred to the variation in obscuration: e.g., both short transition timescale and dramatic optical/infrared variation of CL ANG are inconsistent with the variable obscuration (LaMassa et al. \citeyearpar{LaMassa2015}; MacLeod et al. \citeyear{MacLeod2016}; Sheng et al. \citeyear{Sheng2017}; Gezari et al. \citeyear{Gezari2017}; Wang et al. \citeyear{WangJ2018}); low level polarization of CL AGN suggests the disappearance of the broad emission lines cannot be attributed to dust obscuration (Hutsemeker et al. \citeyear{Husemkerst2017},\citeyear{Husemkerst2019}). 

Even though it is widely suggested that some instabilities in the accretion disk are responsible for the CL mechanism. The driving mechanism itself has pushed standard viscous accretion disc models into a crisis (Lawrence \citeyear{Lawrence2018}), for the accretion theory failed to predict the proper CL transition timescale. Several alternative models have been proposed to explain the CL phenomena, such as X-binary-like accretion (Ruan et al. \citeyear{Run2019}); magnetically elevated accretion (Dexter \& Begelman \citeyear{Dexter2019}); narrow transition zone between the standard disk and inner advection dominated accretion flow (\'{S}niegowska et al. \citeyear{Sniegowska2019}).  

While studying the CL mechanism of AGNs is of particular importance for understanding the structure and physics of accretion disc, CL AGNs also provide an ideal case of investigating the connection between AGNs and their host galaxies such as studying the $M_{BH}-\sigma_{*}$ (Gezari et al. \citeyear{Gezari2017}) or the host of a bright AGN whose starlight is overwhelmed by luminous central engines. 
Furthermore, the timescale and frequency of the transition could help to restrict the lifetimes of AGNs (Martini \& Schneider \citeyear{Martini2003}). So an extensive sample of these CL AGNs is essential addressing above questions.

So far more than 50 CL AGNs have been reported. There are several notable works for finding CL AGNs. For example, mining the archival spectra in the Sloan Digital Sky Survey (SDSS) among quasars with repeated spectra, 2 and 10 new CL AGNs have been identified by Ruan et al. (\citeyear{Ruan2016}) and MacLeod et al. (\citeyear{MacLeod2016}), respectively. Also, Yang et al. (\citeyear{Yang2018}) identified 21 new CL AGNs through similar a method. CL AGNs may be surprisingly common. Recently, Rumbaugth et al (2018) find that extremely variable ($\Delta g>$1) AGN account for 30$\sim$50\% of quasars. Meanwhile, MacLeod et al. \citeyear{MacLeod2019} reported 16 new CL AGNs, yielding a confirmation rate of $\sim$20\%, based on photometric variation selection ($\Delta g>$1 mag and $\Delta r >$0.5) followed by new spectroscopic observations. 

Besides the photometric variation selection method as MacLeod et al. (\citeyear{MacLeod2019}) performed, mid-infrared (MIR) monitoring also provides a powerful tool for finding CL AGNs candidates, and for investigating their physical processes. In our previous work, we investigated 10 CL AGNs (Sheng et al. \citeyear{Sheng2017}) using the MIR multi-photometric monitoring data from Wide-field Infrared Survey Explorer (\emph{WISE}; Wright et al. \citeyear{Wright2010}) and the Near-Earth Object WISE Reactivation mission (\emph{NEOWISE-R}; Mainzer et al. \citeyear{Mainzer2014}). We found all the 10 CL AGNs have strong ($>$ 0.4 mag) variability in MIR bands, and the sources have a transition from Type~1 to Type~2, accompanied by changing from AGN-like MIR-color (\emph{W1}-\emph{W2}$>$0.8) to galaxy-like MIR-color (\emph{W1}-\emph{W2}$<$0.5), vice versa. Meanwhile, the large variability amplitude supports the scheme of dramatic change in the accretion rate. Recently, with large monotonic variation in either \emph{W1} or \emph{W2}, Stern et al. (\citeyear{Stern2018}) reported a new CL AGN J105203.55+151929.5. Assef et al. (\citeyear{Assef2018}) presented two catalogs of AGNs selected from WISE's AllWISE database and found one of the highest variability candidates (WISEA J142846.71+172353.1) was a CL AGN.

Following our previous work, here we propose a method of selecting candidates of CL AGNs based on MIR variation and color change. With this method, we find a large sample of CL AGNs candidates. We followed seven candidates which are bright and proper for observation. Finally, 6 new CL AGNs are identified, and the other one is not robust to be ``turn-off" due to low spectral quality. The outline of this Letter is as follows. In \S\ref{Dt}, we describe the selection method, and the data/spectra used in this study. In \S\ref{dis}, we present the results and make simple discussion. Then we come to a summary in \S\ref{C}. We adopt a flat $\Lambda CDM$ cosmology with $H_0=70~\rm km\;s^{-1} Mpc^{-1}$ and $\Omega_m=0.27$.

\section{Data}\label{Dt}

Our selection criteria aimed at finding good candidates of CL AGNs which already have been spectroscopically identified as AGN with archival Sloan Digital Sky Survey (SDSS). We start from the catalog of all SDSS spectra named as ``\emph{specObj-dr14}". We cull the sources which classified as `QSO' and the redshift is below 0.5 (because we'd like to monitor both $\rm H\beta$ and $\rm H\alpha$). Using the Tool for OPerations on Catalogues And Tables (TOPCAT, Taylor \citeyear{Taylor2005}), we get 6,3680 sources. We also match the DR7, DR12 and Dr14 quasar catalogs (Schneider et al. \citeyear{Schneider2010}; P{\^a}ris et al. \citeyear{P2017},\citeyear{P2018}), and find 307 sources not included in the ``\emph{specObj-dr14}" catalog but satisfied the redshift selection criteria. So we have a sample of 6,4167 sources. We download the multi-epoch photometric data of \emph{WISE} and \emph{NEOWISE} from  NASA/IPAC Infrared Science Archive for the whole sample and extracted mid-IR \emph{W1} (3.4 $\mu m$) and \emph{W2} (4.6 $\mu m$) lightcurves.

As mentioned above, in our previous work we have found CL AGNs can have strong ($>$ 0.4 mag) variability in MIR light curves and color changing along with the type transition.
When the CL AGNs turn from Type~1 to Type~2, their MIR color likely change from AGN-like (\emph{W1}-\emph{W2}$>$0.8) to galaxy-like (\emph{W1}-\emph{W2}$<$0.5), vice versa (Stern et al. \citeyear{Stern2012}; Yan et al. \citeyear{YL2013}; Sheng et al. \citeyear{Sheng2017}). Following our previous results, we constrain that the CLQ candidates should have both significant MIR variation and color transition. So the sources are screened according to the following two criteria: (1) Infrared variability amplitude is larger than 0.4 mag and at more than 5$\sigma$ significance in either \emph{W1} or \emph{W2} bands; (2) the sources whose maximum value of the color (\emph{W1-W2}) below 0.4 or the minimum value above 0.8 are abandoned. After visual vetting of the MIR light curves, we finally select $\sim$300 candidates which have a monotonic decrease tendency in \emph{W1} and \emph{W2} bands (Sheng et al. 2019 in prep.). We took optical spectra of 7 candidates which are bright and proper for observation from our sample: J1252+5918, J1307+4506, J1317+1024, J1428+1723, J1549+1121, J1627+5419, J1713+2736 (full name see column 1 in Table \ref{t1}) in June of 2018.

\subsection{Mid-infrared Light curves}

\begin{figure*}
	\figurenum{1}
	\plotone{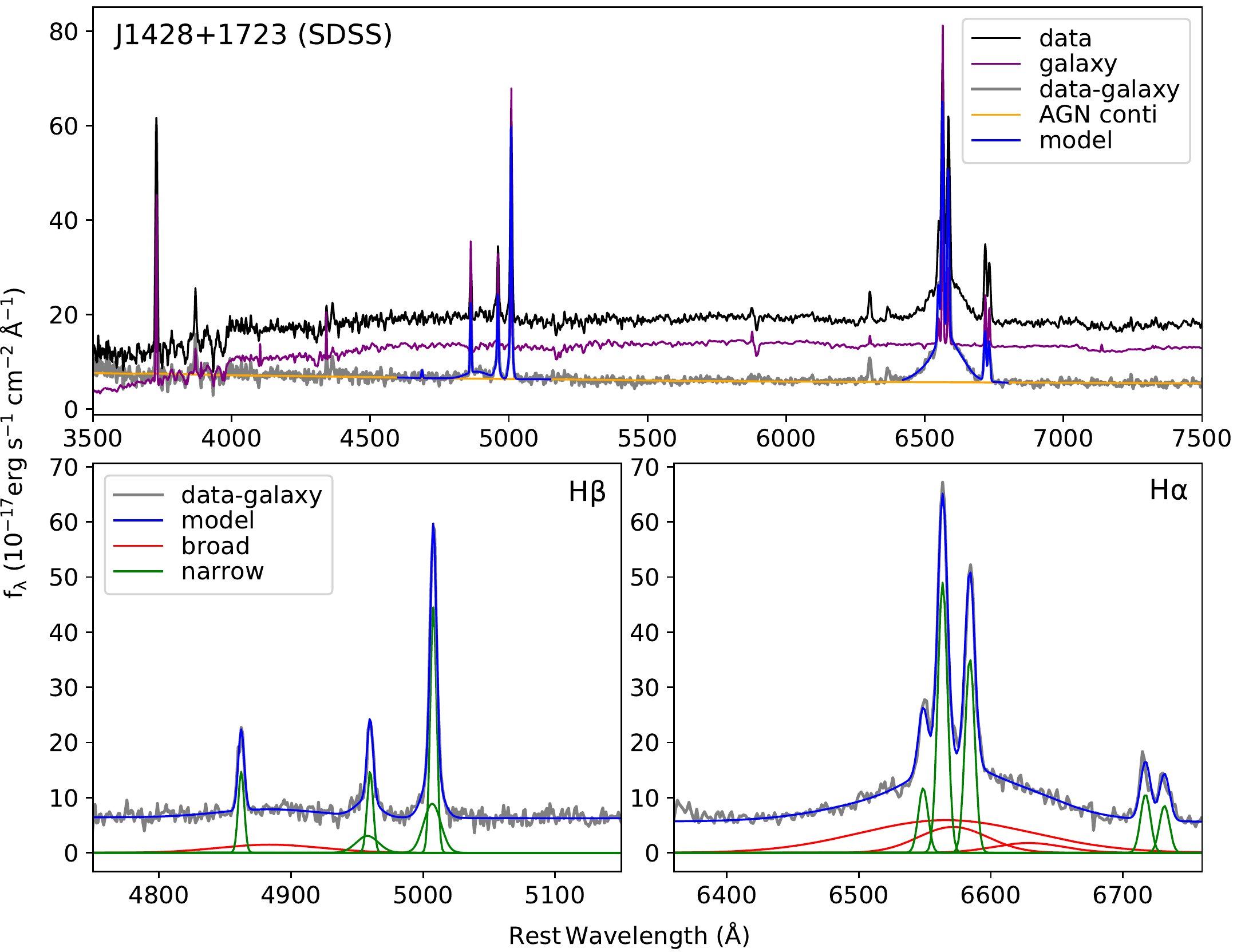}
	\caption{An example of spectral decomposition. In the top panel, the original SDSS spectrum of J1428+1723 is plotted in black. The host galaxy, AGN component and continuum are in purple, grey and orange, respectively. The model for AGN component is in blue. In the bottom panel, we show detailed components of the $\rm H\beta$ and $\rm H\alpha$. The broad components are in red while the narrow ones are in green.}
	\label{fig:1}
\end{figure*}

The \emph{WISE} mission surveyed the entire sky in four infrared wavelengths centered 3.4, 4.6, 12, and 22$\mu$m (denoted  \emph{W1}, \emph{W2}, \emph{W3}, and \emph{W4}) from January to September 2010, until its cryogen used to cool the W3 and W4 channels were depleted. After an extended four months of Post-Cryogenic Mission, it was then placed into hibernation. On 2013 October 3 it is reactivated as \emph{NEOWISE-R}, surveying the sky at \emph{W1} and \emph{W2} (Mainzer et al. \citeyear{Mainzer2014}).  So there is a gap between \emph{WISE} and \emph{NEOWISE} datum. With a polar orbit, \emph{WISE/NEOWISE} scan the entire sky every 6 months and provide $\sim$12 observations per year for the most sources. Following our previous work (Jiang et al. \citeyear{Jiang2012},\citeyear{Jiang2016}; Sheng et al. \citeyear{Sheng2017}), we removed bad data points with poor image quality (``qi\_fact"$<$1), a small separation to South Atlantic Anomaly (``SAA"$<$5) and flagged moon masking (``moon mask"=1), then binned the data every half year using median value. The optical and MIR light curves are presented in the upper left panel of sub-figure of Figure~\ref{fig:1}.

\subsection{Optical spectra}
All the 7 candidates have archive \emph{SDSS} optical spectrum. We acquired another spectrum of J1307+4506, J1428+1723, and J1627+5419, using the Double Spectrograph (DBSP) of the Hale 5m telescope (P200) at the Palomar Observatory on UT 2018 June 06. 
We used a 600/4000 grating for the blue arm and a 316/7500 grating for the red arm, and a D55 dichroic was selected. The sources were observed with two separated 600s exposure through a 1" slit. For the rest candidates J1252+5918, J1317+1024, J1549+1121 and J1713+2736, the spectra were obtained using the Kast double spectrograph at 3m Shane telescope of Lick Observatory on UT 2018 June 11 and 12. We configured the instrument with 1".5 slit, the D57 dichroic, 600/4310 grating on the blue arm and 600/7500 grating on the red arm. The exposure of J1317+1024 was split into two 900s exposure, while others were two 600s. All the data were reduced following the standard IRAF routine and the spectra were calibrated using the standard star \emph{feige56} obtained on the same night. The reduced spectra are plotted in the right hand of Figure \ref{fig:2}. The P200 or Shane DBSP spectrum are scaled to the early SDSS epoch assuming the constant [OIII] emission lines. Using the Python QSO fitting code (PyQSOFit; Guo et al. \citeyear{Guo2018}; Shen et al. \citeyear{Shen2019}), we made decomposition for each source to measure the basic properties such as the luminosity of 5100$\rm \AA$, $\rm H\alpha$ and $\rm H\beta$. An example of decomposition is plotted in Figure \ref{fig:1}. The results are listed in Table \ref{t1}. 

\section{Results and discussion}\label{dis}
Our selection criteria constrain the MIR light curves of candidates should have a feature of large decrease ($>$0.4 mag) tendency along with the MIR color changing from AGN-like to galaxy-like. So all the 7 sources have very similar variation behavior in MIR bands. With the DBSP spectrum decomposition, we confirmed that J1252+5918, J1307+4506, J1317+1024 and J1627+5419 have turned into Type 1.9 due to their disappearance of broad $\rm H\beta$ emission lines, while J1428+1723 even turned into Type 2 because no broad $\rm H\alpha$ component presented in its DBSP spectrum. For J1713+2736, it turned into Type 1.8 because it still has a weak broad $\rm H\beta$ component. For the other one J1549+1121, we also mark it as Type 1.9 AGN because it showed none broad $\rm H\beta$ but weak broad $\rm H\beta$ emission line. However, suffering the low quality spectrum, it is not robust to confirm its totally disappearance of the broad $\rm H\alpha$ component. We listed the detailed type transition of the 7 sources in column 11 and 12 of Table \ref{t1}.

According to the time interval between SDSS and DBSP spectra, 4 AGNs: J1252+5918, J1307+4506, J1317+1024 and J1713+2736 transited from Type 1 to Type 1.9/2 during the timescale of 14$\sim$16 years. However, considering their decreasing tendency of the MIR light curves and color change, the CL transition timescale is likely 4$\sim$6 years in the observational frame, because they kept AGN-like color at MJD$\sim$55500, and turned into galaxy-like around MJD$\sim$57500. The similar transition timescale also presented in J1627+5419, due to its Type~1 like SDSS spectrum is taken at 2013 (MJD=56485) and the Type 1.8 like DBSP spectrum acquired in 2018 (MJD=58276), along with the MIR color (\emph{W1}$-$\emph{W2}) changed from 0.8 to 0.2. 
For J1428+1723, it shows Type 1.8 like feature in its SDSS spectrum which is acquired in 2008, but shows Type 2 like in recent DBSP spectrum. Considering it turned from AGN-like color (\emph{W1}$-$\emph{W2}=1.0) at MJD=55500 into galaxy-like color (\emph{W1}$-$\emph{W2}=0.6) at MJD=57500, the most significant transition might also happen during recent the $\sim$5 years.

While the mechanism of their CL phenomenon is not fully understood, at least 6 of 7 sources have almost the very similar MIR behavior (large decrease and color change in short timescale) to the 10 CL AGNs in our previous work (Sheng et al. \citeyear{Sheng2017}). In that work, we found both the large MIR variability amplitude and the variation timescale of CL AGNs were against the scenario of varying obscuration. And we preferred the scheme of dramatic change in the accretion rate (Sheng et al. \citeyear{Sheng2017}), which is also favored by the most CL studies (e.g. LaMassa et al. \citeyear{LaMassa2015}; Runnoe et al. \citeyear{Runnoe2016};  Ruan et al. \citeyear{Ruan2016}; Macleod et al. \citeyear{MacLeod2016}; Gezari et al. \citeyear{Gezari2017}, Yang et al. \citeyear{Yang2018}; Stern et al. \citeyear{Stern2018}). So the MIR variation is helpful for investigating the mechanism of CL AGNs.

As mentioned above, our CL AGN selection method is based on large decrease ($>$0.4 mag) tendency in the MIR light curve accompanied by the MIR color changing from AGN-like to galaxy-like. This method is benefit from the MIR emission of an AGN is dominated by the dust torus, which provides AGNs with a characteristic red MIR colors and allows a simple \emph{W1}$-$\emph{W2} color cut robustly differentiate AGNs from stars and inactive galaxies (Richard et al. \citeyear{Richard2006}; Stern et al. \citeyear{Stern2012}; Assef et al. \citeyear{Assef2013};). High confirmation rate suggested that it is a very efficient method to select CL AGNs. More detailed discussion along with the sample analysis will be presented in Sheng et al 2019 (in prep.). Moreover, our method can also be applied to select bulks of turn-on CL AGNs form galaxy catalogs, just alternatively constraining the \emph{W1}$-$\emph{W2} changing from galaxy-like to AGN-like.

\section{Summary}\label{C}
We present initial result from a systematic search for CL ANGs based on combining the MIR variation and color change. 
We started from archival quasar spectrum catalog of Sloan Digital Sky Survey (SDSS) and selected a large sample of changing look candidates. 
The sources in the sample should have significant MIR variation ($>$0.4 mag) and change from AGN-like color into galaxy-like color at the same time. We followed 7 candidates from our sample which are bright and proper for observation. And we identified 6 new turn-off CL AGNs, while the other one is not so robust due to the low quality of the spectrum. This work suggested that our selection method is likely to be an effective one to find CL AGNs. A large sample size of CL AGNs can provide us an opportunity to statistically and detailedly investigate the properties of CL ANGs. We will perform a statistical analysis of our sample (Sheng et al. 2019 in prep.) and try to identify more CL AGNs.

\section*{Acknowledgments}
We acknowledge the anonymous referee for valuable comments that helped to improve the Letter. This research uses data obtained through the Telescope Access Program (TAP), which has been funded by the National Astronomical Observatories, Chinese Academy of Sciences, and the Special Fund for Astronomy from the Ministry of Finance. Observations obtained with the Hale Telescope at Palomar Observatory were obtained as part of an agreement between the National Astronomical Observatories, Chinese Academy of Sciences, and the California Institute of Technology. This project is supported by National Basic Research Program of China (grant No. 2015CB857005), the NSFC through NSFC-11233002, NSFC-11421303 and NSFC-116203021. This publication makes use of data products from the Wide-field Infrared Survey Explorer, which is a joint project of the University of California, Los Angeles, and the Jet Propulsion Laboratory/California Institute of Technology, funded by the National Aeronautics and Space Administration. This publication also makes use of data products from \emph{NEOWISE}, which is a project of the Jet Propulsion Laboratory/California Institute of Technology, funded by the Planetary Science Division of the National Aeronautics and Space Administration. This research made use of Astropy, a community-developed core Python package for Astronomy (Astropy Collaboration, \citeyear{Astropy2013}).

\begin{figure*}
	\figurenum{2}
	\epsscale{0.9}
	\plotone{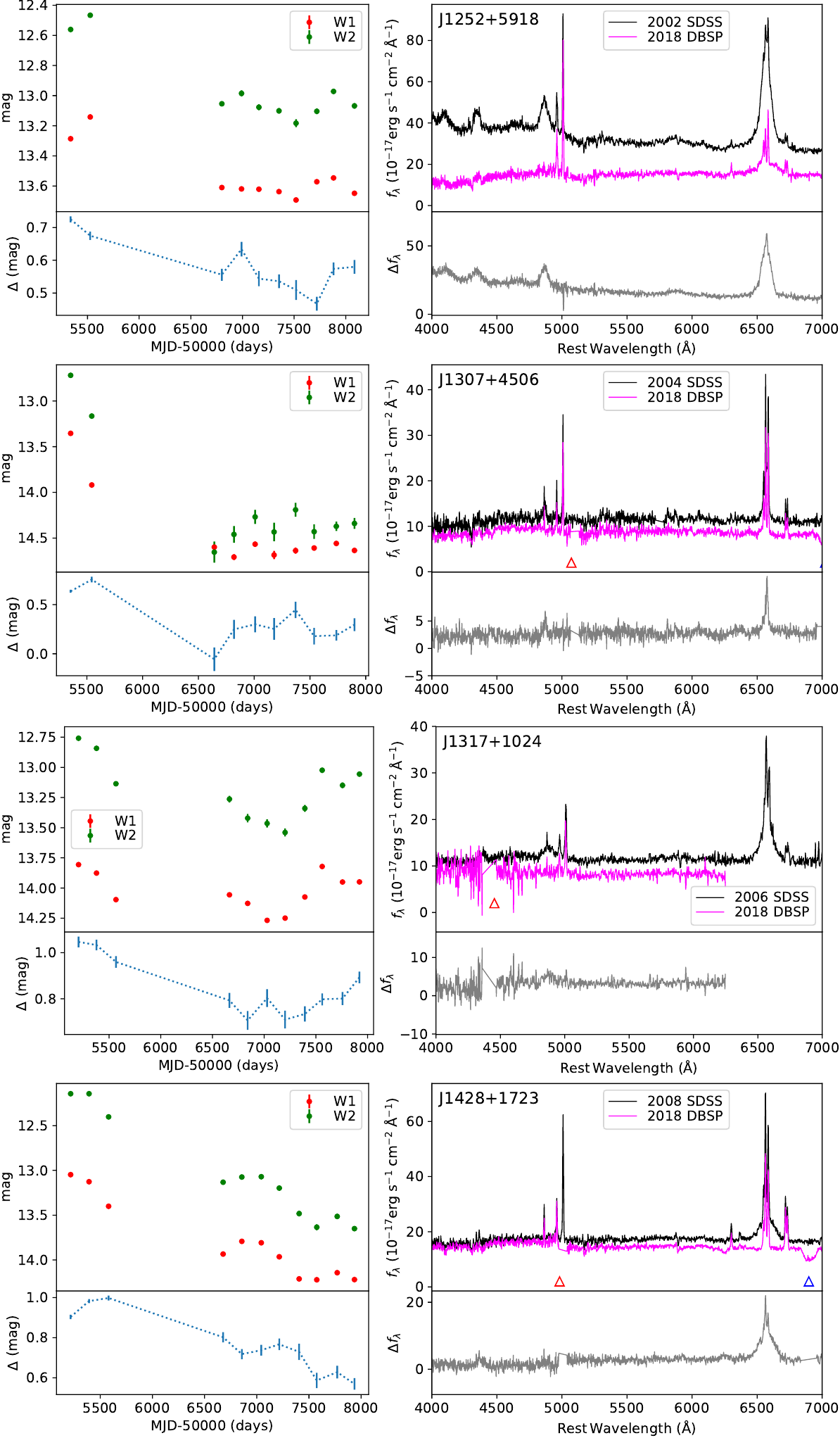}
	\caption{Mid-infrared light curves and spectra. In each figure, the upper left panel presents the mid-infrared light curves. The red and green dots respectively represent the median value of \emph{W1} and \emph{W2} bands, while the corresponding y-error bars are the propagation error. And the lower left panel presents the color of \emph{W1}-\emph{W2}. In the right panel, the previous \emph{SDSS} spectrum is plotted in black while the recent DBSP spectrum is in magenta. And the lower right panel shows the difference of two spectra. The red triangle marks position of the dichroic while the blue one marks uncorrected telluric.}
	\label{fig:2}
\end{figure*}
\renewcommand{\thefigure}{\arabic{figure}}
\addtocounter{figure}{1}
\begin{figure*}
	\centering
	\epsscale{0.9}
	\plotone{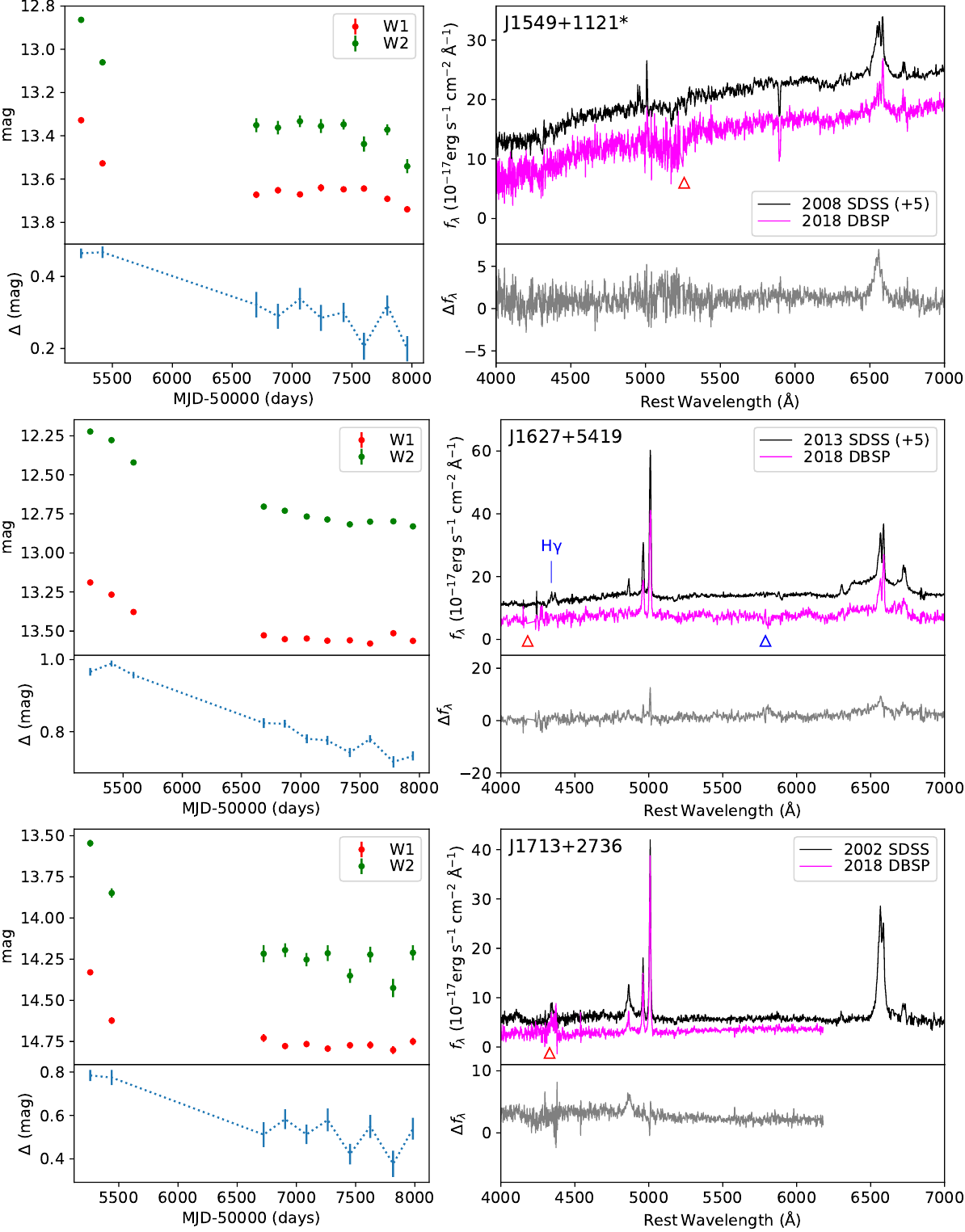}
	\caption{(continued)}
	
\end{figure*}
\begin{table*}
	\caption{\label{t1}Information of 7 AGNs}
	\centering
	\resizebox{\textwidth}{!}{
	\begin{tabular}{lccccccccccc}
		\toprule
		(1)&(2)&(3)&(4)&(5)&(6)&(7)&(8)&(9)&(10)&(11)&(12)\\
		Name  & redshift &Max$\Delta W1$ &Max$\Delta W2$ & $log L_{5100,1}$&$log L_{5100,2}$ &$L_{\rm H\beta,1}$&$L_{\rm H\beta,2}$& $L_{\rm H\alpha,1}$ & $L_{\rm H\alpha,2}$ &T1&T2\\
		      &          &(mag) &(mag)& (\lum)&(\lum)& ($10^{41}$\lum)&($10^{41}$\lum)& ($10^{41}$\lum)&($10^{41}$\lum)& & \\
		\midrule

	J125258.72+591832.7&0.124&$0.55\pm$0.02&$0.72\pm$0.03&43.753$\pm$0.003  &43.065$\pm$0.008 & 6.900$\pm$0.003 &0                   &23.721$\pm$0.084 &3.886$\pm$0.120 & 1 &1.9\\
	J130716.99+450645.3&0.084&$1.35\pm$0.03&$1.93\pm$0.12&42.520$\pm$0.006  &42.223$\pm$0.006 & 0.202$\pm$0.007 &0                   & 0.663$\pm$0.016 &0.060$\pm$0.004 &1.8&1.9\\
	J131737.93+102427.7&0.280&$0.46\pm$0.02&$0.78\pm$0.04 &43.927$\pm$0.009 &43.654$\pm$0.008 & 9.562$\pm$0.521 &0                   &32.587$\pm$0.276 & -              & 1 &1.9\\
	J142846.71+172353.1&0.104&$1.17\pm$0.02&$1.51\pm$0.03 &42.931$\pm$0.006 &42.307$\pm$0.008 & 0.402$\pm$0.045 &0                   & 3.753$\pm$0.148 &0               &1.8& 2 \\
	J154953.60+112148.3$^{*}$&0.084&$0.41\pm$0.02&$0.68\pm$0.03&42.378$\pm$0.004 &42.001$\pm$0.031 & 0.024$\pm$0.011 &0              & 1.072$\pm$0.012 &0.303$\pm$0.031 &1.8&1.9\\
	J162752.18+541912.5&0.316&$0.39\pm$0.01&$0.61\pm$0.01&43.817$\pm$0.002 &43.627$\pm$0.004 & 9.017$\pm$0.260 &0                   &79.397$\pm$7.289 &28.057$\pm$1.931& 1 &1.9\\
	J171353.85+273626.8&0.298&$0.47\pm$0.03&$0.88\pm$0.06&43.902$\pm$0.005 &43.209$\pm$0.077 &15.507$\pm$0.569 &3.015$\pm$0.340     &21.879$\pm$0.727 & -              & 1 &1.8\\
		\bottomrule
	\end{tabular}}
	\tablecomments{In column 1, J154953.60+112148.3 is marked with a star for its type tansition is not so robust due to the low qulity data. Column 3 presents the maximum of $\Delta W1$, while column 4 lists that of $\Delta W2$. $log L_{5100,1}$, $L_{\rm H \beta,1}$ and $L_{\rm H \alpha,1}$ represent the luminosity of 5100$\rm\AA$ (in base-10 logarithm), $\rm H\beta$, $\rm H\alpha$ measured from the previous SDSS spectrum, while the $log L_{5100,2}$, $L_{\rm H \beta,2}$ and $L_{\rm H \alpha,2}$ measured from the recent DBSP spectrum. In column 11 and 12, T1 and T2 respectively represent the spectral types of the previous SDSS spectrum and the recent DBSP spectrum. For J1428+1723 and J1713+2736, their Shane DBSP spectra didn't cover the $\rm H\alpha$, so their $L_{\rm H \alpha,2}$ marked with a dash symbol in column 10.}
\end{table*}


\begin{thebibliography}{}

\bibitem[Assef et al.(2013)]{Assef2013} Assef, R.~J., Stern, D., Kochanek, C.~S., et al.\ 2013, \apj, 772, 26.

\bibitem[Assef et al.(2018)]{Assef2018} Assef, R.~J., Prieto, J.~L., Stern, D., et al.\ 2018, \apj, 866, 26.

\bibitem[Astropy Collaboration et al.(2013)]{Astropy2013} Astropy Collaboration, Robitaille, T.~P., Tollerud, E.~J., et al.\ 2013, \aap, 558, A33 

\bibitem[Cohen et al.(1986)]{Cohen1986} Cohen, R.~D., Puetter, R.~C., Rudy, R.~J., Ake, T.~B., \& Foltz, C.~B.\ 1986, \apj, 311, 135 

\bibitem[Denney et al.(2014)]{Denney2014} Denney, K.~D., De Rosa, G., Croxall, K., et al.\ 2014, \apj, 796, 134 

\bibitem[Dexter, \& Begelman(2019)]{Dexter2019} Dexter, J., \& Begelman, M.~C.\ 2019, \mnras, 483, L17.

\bibitem[Gezari et al.(2017)]{Gezari2017} Gezari, S., Hung, T., Cenko, S.~B., et al.\ 2017, \apj, 835, 144 

\bibitem[Guo et al.(2018)]{Guo2018} Guo, H., Shen, Y., \& Wang, S.\ 2018, Astrophysics Source Code Library, ascl:1809.008 

\bibitem[Hutsem{\'e}kers et al.(2017)]{Husemkerst2017} Hutsem{\'e}kers, D., Ag{\'{\i}}s Gonz{\'a}lez, B., Sluse, D., Ramos Almeida, C., \& Acosta Pulido, J.-A.\ 2017, \aap, 604, L3 

\bibitem[Hutsem{\'e}kers et al.(2019)]{Husemkerst2019} Hutsem{\'e}kers, D., Ag{\'\i}s Gonz{\'a}lez, B., Marin, F., et al.\ 2019, arXiv e-prints , arXiv:1904.03914.

\bibitem[Jiang et al.(2012)]{Jiang2012} Jiang, N., Zhou, H.-Y., Ho, L.~C., et al.\ 2012, \apjl, 759, L31 

\bibitem[Jiang et al.(2016)]{Jiang2016} Jiang, N., Dou, L., Wang, T., et al.\ 2016, \apjl, 828, L14  

\bibitem[Jun et al.(2015)]{Jun2015} Jun, H.~D., Stern, D., Graham, M.~J., et al.\ 2015, \apjl, 814, L12 

\bibitem[LaMassa et al.(2015)]{LaMassa2015} LaMassa, S.~M., Cales, S., Moran, E.~C., et al.\ 2015, \apj, 800, 144 

\bibitem[Lawrence(2018)]{Lawrence2018} Lawrence, A.\ 2018, Nature Astronomy, 2, 102 

\bibitem[MacLeod et al.(2016)]{MacLeod2016} MacLeod, C.~L., Ross, N.~P., Lawrence, A., et al.\ 2016, \mnras, 457, 389 

\bibitem[MacLeod et al.(2019)]{MacLeod2019} MacLeod, C.~L., Green, P.~J., Anderson, S.~F., et al.\ 2019, \apj, 874, 8.

\bibitem[Martini \& Schneider(2003)]{Martini2003} Martini, P., \& Schneider, D.~P.\ 2003, \apjl, 597, L109 

\bibitem[Mainzer et al.(2014)]{Mainzer2014} Mainzer, A., Bauer, J., Cutri, R.~M., et al.\ 2014, \apj, 792, 30  

\bibitem[Oknyansky et al.(2019)]{Oknyansky2019} Oknyansky, V.~L., Winkler, H., Tsygankov, S.~S., et al.\ 2019, \mnras, 483, 558 

\bibitem[P{\^a}ris et al.(2017)]{P2017} P{\^a}ris, I., Petitjean, P., Ross, N.~P., et al.\ 2017, \aap, 597, A79 

\bibitem[P{\^a}ris et al.(2018)]{P2018} P{\^a}ris, I., Petitjean, P., Aubourg, {\'E}., et al.\ 2018, \aap, 613, A51.

\bibitem[Richards et al.(2006)]{Richard2006} Richards, G.~T., Lacy, M., Storrie-Lombardi, L.~J., et al.\ 2006, \apjs, 166, 470.

\bibitem[Ruan et al.(2016)]{Ruan2016} Ruan, J.~J., Anderson, S.~F., Cales, S.~L., et al.\ 2016, \apj, 826, 188 

\bibitem[Runco et al.(2016)]{Runco2016} Runco, J.~N., Cosens, M., Bennert, V.~N., et al.\ 2016, \apj, 821, 33 

\bibitem[Ruan et al.(2019)]{Run2019} Ruan, J.~J., Anderson, S.~F., Eracleous, M., et al.\ 2019, arXiv e-prints , arXiv:1903.02553.

\bibitem[Runnoe et al.(2016)]{Runnoe2016} Runnoe, J.~C., Cales, S., Ruan, J.~J., et al.\ 2016, \mnras, 455, 1691 

\bibitem[Schneider et al.(2010)]{Schneider2010} Schneider, D.~P., Richards, G.~T., Hall, P.~B., et al.\ 2010, \aj, 139, 2360.

\bibitem[Shappee et al.(2014)]{Shappee2014} Shappee, B.~J., Prieto, J.~L., Grupe, D., et al.\ 2014, \apj, 788, 48  

\bibitem[Shen et al.(2019)]{Shen2019} Shen, Y., Hall, P.~B., Horne, K., et al.\ 2019, \apjs, 241, 34.

\bibitem[Sheng et al.(2017)]{Sheng2017} Sheng, Z., Wang, T., Jiang, N., et al.\ 2017, \apj, 846, L7.

\bibitem[{\'S}niegowska, \& Czerny(2019)]{Sniegowska2019} {\'S}niegowska, M., \& Czerny, B.\ 2019, arXiv e-prints , arXiv:1904.06767.

\bibitem[Stern et al.(2012)]{Stern2012} Stern, D., Assef, R.~J., Benford, D.~J., et al.\ 2012, \apj, 753, 30   

\bibitem[Stern et al.(2018)]{Stern2018} Stern, D., McKernan, B., Graham, M.~J., et al.\ 2018, \apj, 864, 27.

\bibitem[Taylor(2005)]{Taylor2005} Taylor, M.~B.\ 2005, Astronomical Data Analysis Software and Systems XIV, 347, 29 

\bibitem[Wang et al.(2018)]{WangJ2018} Wang, J., Xu, D.~W., \& Wei, J.~Y.\ 2018, \apj, 858, 49 

\bibitem[Wright et al.(2010)]{Wright2010} Wright, E.~L., Eisenhardt, P.~R.~M., Mainzer, A.~K., et al.\ 2010, \aj, 140, 1868-1881 

\bibitem[Yan et al.(2013)]{YL2013} Yan, L., Donoso, E., Tsai, C.-W., et al.\ 2013, \aj, 145, 55 

\bibitem[Yang et al.(2018)]{Yang2018} Yang, Q., Wu, X.-B., Fan, X., et al.\ 2018, \apj, 862, 109.

\end{thebibliography}
\end{document}